\documentclass{aa}

\usepackage{natbib}

\newcommand{\fuse}{\emph{FUSE}}
\newcommand{\hst}{\emph{HST}}
\newcommand{\stis}{\emph{HST/STIS}}
\newcommand{\iue}{\emph{IUE}}

\begin{document}

\title{Beyond the iron group: heavy metals in hot subdwarfs\thanks{Based on
    observations made with the NASA/ESA Hubble Space Telescope, which is
    operated by the Association of Universities for Research in Astronomy,
    Inc., under NASA contract NAS 5-26555. These observations are associated
    with program \#8635 and \#5319.}
}

\author{S.~J. O'Toole}

\institute{Dr Remeis-Sternwarte, Astronomisches Institut der
  Universit\"at Erlangen-N\"urnberg, Sternwartstr. 7, Bamberg D-96049,
  Germany, \email{otoole@sternwarte.uni-erlangen.de}}

\offprints{Simon O'Toole}

\date{Received / Accepted}

\abstract{We report the discovery of strong photospheric resonance lines of
  Ga\,\textsc{iii}, Ge\,\textsc{iv}, Sn\,\textsc{iv} and
  Pb\,\textsc{iv} in the UV spectra of more than two dozen sdB and sdOB
  stars at temperatures ranging from 22\,000\,K to 40\,000\,K. Lines of other
  heavy elements are also detected, however in these cases more atomic data are
  needed. Based on these discoveries, we present a hypothesis to explain the
  apparent lack of silicon in sdB stars hotter than $\sim$32\,000\,K. The
  existence of triply ionised Ge, Sn, and Pb suggests that rather than silicon
  sinking deep into the photosphere, it is removed from the star in a
  fractionated stellar wind. This hypothesis provides a challenge to diffusion
  models of sdB stars.
}

\titlerunning{Heavy metals in hot subdwarfs}
\authorrunning{S.~J. O'Toole}

\maketitle

\section{Introduction}
\label{sec:intro}

Improving atomic data and access to high-resolution UV spectra through
\hst\ and \fuse\ have allowed abundance analyses of hot, degenerate stars to be
undertaken \citep[e.g.\ ][]{OCM00,BGH03}. We have recently begun a project to
measure the abundances of iron-group elements in hot subdwarf B stars
\citep{OHC03,CFF03}. These stars can be associated with models of hot, helium
core-burning stars with hydrogen layers too thin to sustain nuclear
burning. Analyses of high resolution \iue\ spectra of sdBs revealed peculiar
chemical abundances. In particular, carbon varies greatly from star
to star, nitrogen is essentially solar, while silicon is slightly
underabundant up to $\sim$32\,000\,K, but at hotter temperatures it is very
deficient -- at least 5 dex below the solar value in some cases
\citep[e.g.\ ][]{MontrealV,BHS82}. These chemical pecularities in sdBs are
usually explained by invoking a complex balance of radiative leviation and
gravitational settling combined with mass loss \citep{UB01,MontrealVI},
however, there have only been a few diffusion models published, with varying
success \citep{UB01,CFB97a}.

In this Letter we present the discovery of strong resonance lines of several
heavy elements in sdB (there will be no distinction made between
sdB and sdOB stars here) based on high resolution UV
spectra. Three of the elements discovered (germanium, tin and lead) have the
same electronic configuration as silicon, and we discuss the implications of
this in the context of diffusion theory. A detailed abundance analysis is
currently being carried out (O'Toole et al., in preparation).

\section{Observations}
\label{sec:obs}

Our original observations were made with \stis\ as part of an ongoing project
to measure metal abundances in pulsating and non-pulsating sdBs (proposal
\#8635, PI: U.\ Heber). We obtained data for five targets using the E140M
grating, which in the far UV has a resolving power of $R=45\,800$ and a useful
wavelength range of 1165-1710\,\AA. Final results will be presented in a
future paper, however a preliminary analysis can be found in \citet{OHC03},
while the serendipitous discovery of a white dwarf companion to one of the
stars (Feige 48) using these data has been presented by \citet{OHB04}.

\begin{figure}
\vspace{5.7cm}
\begin{center}
    \includegraphics{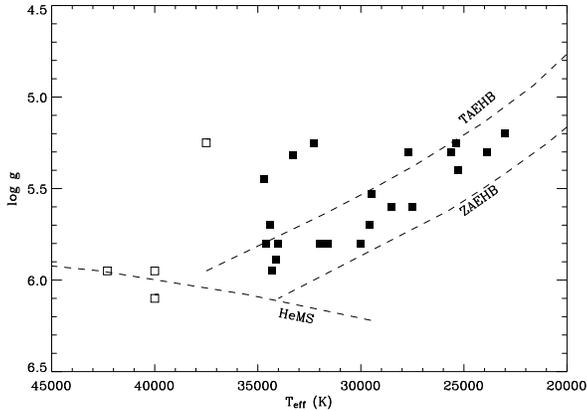}
\end{center}
\caption{$T_{\mathrm{eff}}-\log g$ diagram of the sample stars. Filled squares
  show sdB stars, while open squares show sdOs. Two stars are not shown since
  they do not have temperature determinations.}
\label{fig:teff}
\end{figure}

After our original discovery of heavy element resonance lines in our \stis\
spectra, we searched for any other spectra of sdB/O stars in the \hst\ and
\iue\ archives. This resulted in high-resolution UV \stis\ spectra of Feige 66
and CPD\,$-64^\circ 481$, a \emph{HST/GHRS} spectrum of the sdO Feige 46, and
the \iue\ spectra of 21 more stars,
representing almost the entire range of effective temperatures and surface
gravities for the sdB stars (see Figure \ref{fig:teff}). 
In one case (Feige~110) the detection of the Pb\,\textsc{iv} line at
1313.072\,\AA\ was uncertain, so we looked for the second resonance line at
1028.611\,\AA\ using archival \fuse\ spectra.

\section{Discovery of heavy element resonance lines}
\label{sec:atomsynth}

\begin{figure}
\vspace{10cm}
\begin{center}
    \includegraphics{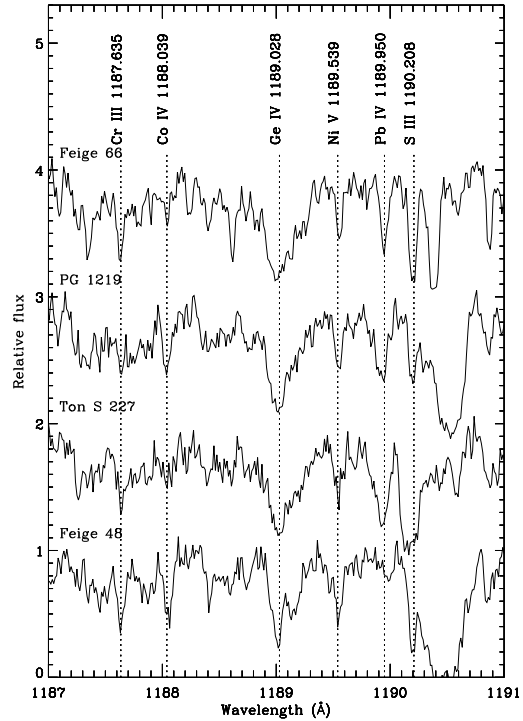}
\end{center}
\caption{\stis\ spectra of 1187-1191\,\AA\ region showing the Ge\,\textsc{iv}
  1189.028\,\AA\ line, as well as the Pb\,\textsc{iv} at 1189.95\,\AA. There
  is no atomic data for the latter. The strong feature around 1190.5\,\AA\ is
  due to interstellar Si\,\textsc{ii}. Several other strong lines are also
  marked.}
\label{fig:gestis}
\end{figure}

The analysis of the UV spectra of sdBs is complicated by the large number of
absorption lines caused by iron-group elements. The wavelengths of these lines
in the Kurucz database in some cases can be quite inaccurate, especially at
bluer wavelengths. This
can mean that some lines in the spectrum appear to be unidentified. While
analysing our \stis\ spectra we noticed that the wavelengths of two very
strong unmatched lines corresponded very well with the resonance doublet of
Ge\,\textsc{iv}. The region around the line at 1189.028\,\AA\ is shown in
Figure \ref{fig:gestis} (the other line is at 1229.840\,\AA). There are no
strong lines near these resonance lines that could lead to an incorrect
identification. Using the line lists of \citet{Morton2003,Morton2000}, we
discovered the resonance doublets of Ga\,\textsc{iii} at 1495.045\,\AA\ and
1534.462\,\AA\ and of Sn\,\textsc{iv} at 1314.539\,\AA\ and 1437.525\,\AA, as
well as one resonance line of Pb\,\textsc{iv} at 1313.072\,\AA\ (the other
line is at 1028.611\,\AA, lies in the wings of Ly$\beta$). The strongest
Sn\,\textsc{iii} resonance line at 1251.387\,\AA\ may also be present, but is
blended with an unknown feature. The three resonance lines of Zr\,\textsc{iv},
as well as Zn\,\textsc{iv} and possibly Cd\,\textsc{iv} are present in at
least Feige 66, however no oscillator strengths for these lines could be found
in the literature. We note
that the only other report of these resonance lines in UV spectra was by
\citet{PSR01}, who discussed the lead, tin and germanium abundances of the SMC
main-sequence B star AV\,304. They compared this star with the Galactic B star
HD\,46328. Both of these stars show the same Ge\,\textsc{iv}, Sn\,\textsc{iv}
and Pb\,\textsc{iv} lines seen in our spectra. It is quite possible that these
lines have been previously overlooked in the spectra of other main sequence O
and B stars, although with many objects rapidly rotating, their identification
might be difficult. This is one clear advantage that hot subdwarfs have: their
rotation velocities are typically less than 5\,km\,$^{-1}$.

The discovery of the resonance lines of these heavy elements encouraged us to
look for subordinate lines. These lines should be much weaker, but may still
be detectable. We used the NIST Atomic Spectra
Database\protect\footnote{see
  \texttt{http://physics.nist.gov/cgi-bin/AtData/lines\_form}} as well as the
line list of \citet{HH95}. In the case of Ge\,\textsc{iv}, there are two
strong lines around 1494.9\,\AA\ and 1500.6\,\AA\ that have listed oscillator
strengths, however because the wavelengths are not very well known, we cannot
claim a certain detection. There are also a couple of strong Ga\,\textsc{iii}
lines with published oscillator strengths, and these are detected at
1267.151\,\AA\ and 1293.446\,\AA. Several subordinate Pb\,\textsc{iv} lines
with reasonably accurate wavelengths are seen, however once again there are no
oscillator strengths available in the literature. The strongest appears to be
the line at 1189.95\,\AA, which is clearly present in the three hottest stars
Feige~66, PG\,1219+534 and Ton~S~227, but weak in Feige~48 (see Figure
\ref{fig:gestis}). Other weaker Pb\,\textsc{iv} are possibly detected at
1343.06\,\AA, 1400.26\,\AA, and 1535.71\,\AA. There are no strong
Sn\,\textsc{iv} lines other than the resonance lines detectable in our spectra.

\begin{figure}
\vspace{10.5cm}
\begin{center}
    \includegraphics{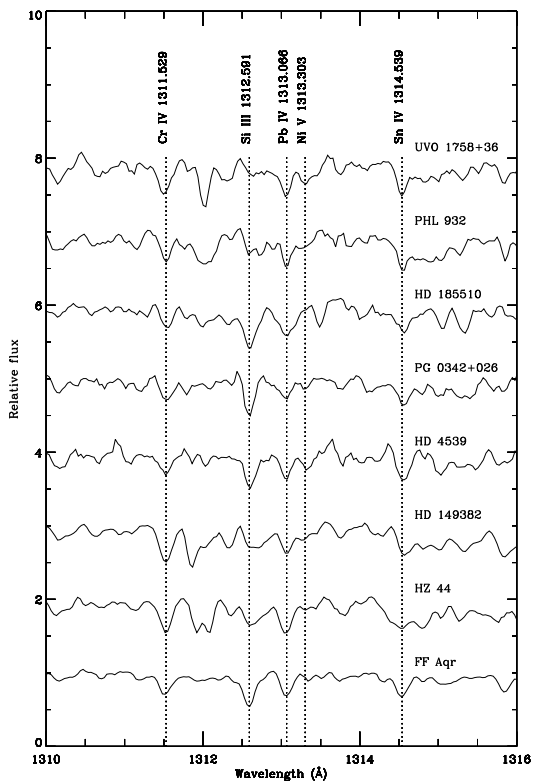}
\end{center}
\caption{\iue\ spectra of the 1311-1315\,\AA\ region showing the
  Sn\,\textsc{iv} and Pb\,\textsc{iv} resonance lines. The
  spectra have been offset for clarity and smoothed with a boxcar filter two
  pixels wide.}
\label{fig:pbiue}
\end{figure}

Our \stis\ spectra are not the only high-resolution UV spectra available
covering this wavelength range. There were a number of sdBs
observed with \iue, and their resolution is good enough to
be able to detect some of the lines we have discovered here, although in some
cases low signal-to-noise makes things difficult. We show a list of all the
stars in our sample in Table \ref{tab:heavy}, and a selection of spectra
around the Pb\,\textsc{iv} and Sn\,\textsc{iv} resonance lines in Figure
\ref{fig:pbiue}. In almost all stars the lines of Ge, Sn, and Pb are
present. Low signal-to-noise is responsible in most cases for uncertainty,
although in some stars the spectra are smeared by rotation (PG\,1605+072,
SB\,290)or tidally-locked binary motion (HW~Vir, PG\,0113+114). The stars in
Table \ref{tab:heavy} are grouped into sdOs at the top, ``cool'' sdBs in the
middle, and ``hot'' sdBs (also known as sdOBs) at the bottom. The cool and hot
sdBs are divided arbitrarily by the appearance of Si\,\textsc{iv} resonance
lines (see Section \ref{sec:disc}). HZ~44 is an unusual object, since it is
hot but shows the same heavy metal lines as cooler sdB stars. This may be due
to convection related to the stars high helium abundance
(log(He/H)$\sim$0.4). The sdO stars SB\,884 and Feige~46 are peculiar and may
not have evolved from the Horizontal Branch. The \iue\ spectrum of the
hydrogen-rich sdO Feige 110
has several bad pixels in the 1310-1315\,\AA, making identification of lines
in this region difficult. We note that the second Pb\,\textsc{iv} resonance
line at 1028.611\,\AA\ may be present in the \fuse\ spectrum of
Feige~110, however it appears to be blended, meaning that the presence of
lead remains uncertain. The case of the Sn\,\textsc{iv} 1437.525\,\AA\ line is
similar. The last column in Table \ref{tab:heavy} shows the binary
status of the stars; the presence of heavy element lines appears to be
independent of binarity. Finally, three of the stars in our sample are
pulsating sdBs, and two of them clearly show the lines discussed, while the
third (PG\,1605+072) is rapidly rotating, making line identification
difficult.

\begin{table}
\caption{List of sdB/sdO stars with high-resolution UV
  spectra, comparing presence of G14 elements. Observations of stars marked
  with an asterisk were made with \stis\ ; all other observations were made
  with \iue. The spectra of HW Vir and PG\,0133+114 are smeared by binary
  motion, while the PG\,1605+072 and SB\,290 spectra are smeared by rotation.}
\label{tab:heavy}
\begin{center}
\begin{tabular}{lccccc}
\hline
Star & Si & Ge & Sn & Pb & binary\\
\hline
HZ 44 & Y & Y & Y & Y & ? \\
SB\,884 & Y & Y & ? & Y & ? \\
Feige 46\rlap{$^*$} & Y & ? & ? & Y & ?\\
Feige 110 & N & Y & ? & ? & ? \\
SB 290 & Y & ? & ? & ? & ? \\
HD\,4539 & Y & Y & Y & Y & N \\
HD\,171858 & Y & Y & Y & Y & Y \\
HD\,185510 & Y & ? & Y & Y & Y \\
HD\,205805 & Y & Y & Y & Y & N \\
FF Aqr & Y & Y & Y & Y & Y \\
HW Vir & Y & ? & ? & ? & Y \\
Feige 48\rlap{$^*$} & Y & Y & Y & Y & Y \\
Feige 65 & Y & ? & Y & Y & N \\
CPD$-64^\circ481$\rlap{$^*$} & Y &  Y & Y & Y & ? \\
PG\,0133+114 & Y & ? & ? & ? & Y \\
PG\,0342+026 & Y & Y & Y & Y & N \\
PG\,1605+072\rlap{$^*$} & Y & Y & ? & ? & ?\\
PG\,1032+406 & N & Y & Y & Y & Y \\
PG\,1104+243 & N & Y & Y & Y & Y \\
PG\,1219+534\rlap{$^*$} & N & Y & Y & Y & ? \\
PG\,1352-022 & N & Y & ? & Y & ? \\
HD\,149382 & N & Y & Y & Y & N \\
Feige 66\rlap{$^*$} & N & Y & Y & Y & N \\
Ton S 227\rlap{$^*$} & N & Y & Y & Y & ? \\
BD$+34^\circ1543$ & N & Y & ? & Y & Y \\
PHL\,932 & N & Y & Y & Y & N \\
UVO\,1758+36 & N & Y & Y & Y & N \\
\hline
\end{tabular}
\end{center}
\end{table}

\section{Implications}
\label{sec:disc}

The effects of diffusion hamper any meaningful discussion about the origin of
the heavy elements discussed in this Letter (i.e.\ whether they are formed by
the $r$- or $s$-process). Clearly they can not be formed in these
hot subdwarfs, but their presence can, however, be used to shed some light
on some of the other chemical peculiarities found in these stars.

It has long been know that silicon in sdB stars show peculiar
behaviour. Above a certain temperature ($\sim$32\,000\,K) there is almost no
silicon left in the upper photosphere. Previous studies suggested that above
this temperature silicon might exist in a noble gas configuration
\citep[e.g. ][]{BHS82}, and that it should sink deep into the
photosphere since these atoms would absorb little radiative flux, and hence
momentum. Calculations by \citet{MontrealIV} and \citet{MontrealVI} have
found, however, that diffusion alone cannot account for the large silicon
underabundances, and that a competing particle transport mechanism, i.e.\ a
weak stellar wind, must be present. The discovery of strong Ge, Sn and Pb
resonance lines can be used to confirm this. Consider that Si, Ge, Sn and Pb
occupy the same column in the periodic table (Group 14, hereafter G14), i.e.\
they each have 4 electrons in their outer shell. (Note that C\,\textsc{iv},
the lowest mass G14 element, has a significantly higher ionisation potential
(64.492\,eV), and is also produced by the triple-$\alpha$ process, meaning
that it is somewhat independent of the other G14 elements). This means that
we can expect their susceptibility to absorb
radiative momentum to be quite similar. In other words, to a first
approximation, the elements' initial abundances and relative masses will be
the only things that affect the outcome of their acceleration by radiative
flux. As a consequence, the velocity of Pb atoms will be seven times lower
than Si atoms, since Si is seven times lighter than lead. Consider that the
escape velocity of a typical sdB is $\sim$1200\,km\,s$^{-1}$; if
$v\mathrm{(Si)}$=2000\,km\,s$^{-1}$ (not an unrealistic value), then
$v\mathrm{(Pb)}$=$\sim$270\,km\,s$^{-1}$,
$v\mathrm{(Sn)}$=$\sim$475\,km\,s$^{-1}$, and
$v\mathrm{(Ge)}$=$\sim$770\,km\,s$^{-1}$. In other words, silicon will
escape the star, but the heavier elements will not.
A similar qualitative argument was put forward by \citet{LPW99} to
explain the abundance peculiarities of the HgMn star $\chi$ Lupi.

An implication of the hypothesis presented here is that this wind must be
fractionated, rather than uniform, since Coulomb coupling is
inefficient in extremely thin winds. \citet{FC97} investigated the effects of
a weak stellar wind on helium in sdBs and found that the observed abundances
can be explained if the mass-loss rates are between about $10^{-14}$ and
$10^{-13} M_\odot$\,yr$^{-1}$. \citet{UB01} found that the CNO elements could
show strong deficiencies as well as enrichments when they considered the
effects of such a wind with these kinds of mass-loss rates. Detecting such a
wind directly would be difficult for most sdBs, but \citet{HMM03} discovered
four luminous objects that show weak H$\alpha$ emission cores, and suggested
that these may be the signature of a stellar wind. Preliminary calculations by
\citet{Vink2004} show that the mass-loss rates for these stars may be as high
as 10$^{-11} M_\odot$. Because none of sdB stellar wind studies mentioned here
have done so, we suggest that future calculations include silicon.

Naively one might expect atoms with similar mass to be affected in the same
way as silicon. For lighter elements this appears to be the case, with
aluminium also at least significantly depleted in hotter sdBs
\citep[e.g.][]{BHS82}. The presence of Ga\,\textsc{iii}, with a similar
ionisation potential, suggests that Al\,\textsc{iii} should be present in
these stars, rather than Al\,\textsc{iv}. For the main ionisation stages of
the other elements (Mg\,\textsc{iii}, P\,\textsc{iii/iv}, S\,\textsc{iii/iv},
Ar\,\textsc{iii}) in Period 3, however, most flux is absorbed by lines below
the Lyman limit, where very little is emitted by sdB stars. As a consequence,
these elements will not be as strongly driven as aluminium or silicon, meaning
they should be present in the spectra of sdB stars. This is the case for all
species except Mg\,\textsc{iii}, which has very few strong lines at observable
wavelengths.

\section{Summary}
\label{sec:summary}

We have discovered strong resonance lines of lead, tin, germanium
and gallium in many sdB stars. The presence of these lines is independent of
binarity, and they are present in both pulsating and non-pulsating sdBs. Other
heavy elements may also be present, however lists of laboratory wavelengths
and oscillator strengths are missing or incomplete. The lines should also be
present main sequence O and B stars. We suggest that the
presence of heavy Group 14 elements, but the near absence of silicon in the
hotter sdBs, implies that rather than sinking deeping into the photosphere,
silicon may have been carried out of the star by a fractionated stellar
wind. This hypothesis is supported by measurements of aluminium resonance
lines, which also show strong depletion in hotter sdB stars. These discoveries
provide a challenge for future diffusion and stellar wind models of hot
subdwarfs.

\begin{acknowledgements}

We are thankful to Norbert Przybilla for many helpful comments, as well as Uli
Heber and Klaus Unglaub for useful discussions.  SJOT is supported by the
Deutsches Zentrum f\"ur Luft- und Raumfahrt (DLR) through grant no.\
50-OR-0202.
\end{acknowledgements}

\bibliographystyle{aa}

\begin{thebibliography}{19}
\expandafter\ifx\csname natexlab\endcsname\relax\def\natexlab#1{#1}\fi

\bibitem[{{Barstow} {et~al.}(2003){Barstow}, {Good}, {Holberg}, {Hubeny},
  {Bannister}, {Bruhweiler}, {Burleigh}, \& {Napiwotzki}}]{BGH03}
{Barstow}, M.~A., {Good}, S.~A., {Holberg}, J.~B., {et~al.} 2003, MNRAS, 341,
  870

\bibitem[{{Baschek} {et~al.}(1982){Baschek}, {Hoeflich}, \& {Scholz}}]{BHS82}
{Baschek}, B., {Hoeflich}, P., \& {Scholz}, M. 1982, A\&A, 112, 76

\bibitem[{{Bergeron} {et~al.}(1988){Bergeron}, {Wesemael}, {Michaud}, \&
  {Fontaine}}]{MontrealVI}
{Bergeron}, P., {Wesemael}, F., {Michaud}, G., \& {Fontaine}, G. 1988, ApJ,
  332, 964

\bibitem[{{Charpinet} {et~al.}(1997){Charpinet}, {Fontaine}, {Brassard},
  {Chayer}, {Rogers}, {Iglesias}, \& {Dorman}}]{CFB97a}
{Charpinet}, S., {Fontaine}, G., {Brassard}, P., {et~al.} 1997, ApJ, 483, L123

\bibitem[{{Chayer} {et~al.}(2003){Chayer}, {Fontaine}, {Fontaine},
  {Lamontagne}, {Wesemael}, {Dupuis}, {Heber}, {Napiwotzki}, \&
  {Moehler}}]{CFF03}
{Chayer}, P., {Fontaine}, G., {Fontaine}, M., {et~al.} 2003, in {Extreme
  Horizontal Branch stars and related objects}, ed. P.~{Maxted} (Kluwer), in
  press

\bibitem[{{Fontaine} \& {Chayer}(1997)}]{FC97}
{Fontaine}, G. \& {Chayer}, P. 1997, in The Third Conference on Faint Blue
  Stars, 169

\bibitem[{{Heber} {et~al.}(2003){Heber}, {Maxted}, {Marsh}, {Knigge}, \&
  {Drew}}]{HMM03}
{Heber}, U., {Maxted}, P.~F.~L., {Marsh}, T.~R., {Knigge}, C., \& {Drew}, J.~E.
  2003, in NATO ASIB Proc. 105: White Dwarfs, 109

\bibitem[{{Hirata} \& {Horaguchi}(1995)}]{HH95}
{Hirata}, R. \& {Horaguchi}, T. 1995, VizieR On-line Data Catalog: VI/69

\bibitem[{{Lamontagne} {et~al.}(1987){Lamontagne}, {Wesemael}, \&
  {Fontaine}}]{MontrealV}
{Lamontagne}, R., {Wesemael}, F., \& {Fontaine}, G. 1987, ApJ, 318, 844

\bibitem[{{Leckrone} {et~al.}(1999){Leckrone}, {Proffitt}, {Wahlgren},
  {Johansson}, \& {Brage}}]{LPW99}
{Leckrone}, D.~S., {Proffitt}, C.~R., {Wahlgren}, G.~M., {Johansson}, S.~G., \&
  {Brage}, T. 1999, AJ, 117, 1454

\bibitem[{{Michaud} {et~al.}(1985){Michaud}, {Bergeron}, {Wesemael}, \&
  {Fontaine}}]{MontrealIV}
{Michaud}, G., {Bergeron}, P., {Wesemael}, F., \& {Fontaine}, G. 1985, ApJ,
  299, 741

\bibitem[{{Morton}(2000)}]{Morton2000}
{Morton}, D.~C. 2000, ApJS, 130, 403

\bibitem[{{Morton}(2003)}]{Morton2003}
---. 2003, ApJS, 149, 205

\bibitem[{{Ohl} {et~al.}(2000){Ohl}, {Chayer}, \& {Moos}}]{OCM00}
{Ohl}, R.~G., {Chayer}, P., \& {Moos}, H.~W. 2000, ApJ, 538, L95

\bibitem[{{O'Toole} {et~al.}(2004){O'Toole}, {Heber}, \& {Benjamin}}]{OHB04}
{O'Toole}, S.~J., {Heber}, U., \& {Benjamin}, R.~A. 2004, A\&A, in press

\bibitem[{{O'Toole} {et~al.}(2003){O'Toole}, {Heber}, {Chayer}, {Fontaine},
  {O'Donoghue}, \& {Charpinet}}]{OHC03}
{O'Toole}, S.~J., {Heber}, U., {Chayer}, P., {et~al.} 2003, in {Extreme
  Horizontal Branch stars and related objects}, ed. P.~{Maxted} (Kluwer),
  astro-ph/0309062

\bibitem[{{Proffitt} {et~al.}(2001){Proffitt}, {Sansonetti}, \&
  {Reader}}]{PSR01}
{Proffitt}, C.~R., {Sansonetti}, C.~J., \& {Reader}, J. 2001, ApJ, 557, 320

\bibitem[{{Unglaub} \& {Bues}(2001)}]{UB01}
{Unglaub}, K. \& {Bues}, I. 2001, A\&A, 374, 570

\bibitem[{{Vink}(2004)}]{Vink2004}
{Vink}, J. 2004, in {Extreme Horizontal Branch stars and related objects}, ed.
  P.~{Maxted} (Kluwer), astro-ph/0309011

\end{thebibliography}

\end{document}